\documentclass[usenatbib]{mn2e} 
\usepackage{graphicx}
\usepackage{longtable} 
\def\aj{AJ}
\def\apj{ApJ}
\def\aa{A\&A}
\def\aas{A\&AS}
\def\mnras{MNRAS}

\title[Astrometry of NGC 3766]{Proper motions and membership probabilities of stars in the region of open cluster NGC~3766.\thanks{Based   on
   observations with the MPG/ESO  {\bf 2.2}m, located
   at  La Silla  and  Paranal Observatory,  Chile,  under DDT  programs
   164.O-0561(E) and 077.C-0188(B).}} 

\author[Yadav et al.]
{ R.\ K.\ S.\ Yadav$^{1}$\thanks{E-mail: rkant; devesh; sagar@aries.res.in}
Devesh\ P.\  Sariya$^{1,2}$,  
R.\ Sagar$^{1}$\\
\\ 
$^{1}$Aryabhatta Research Institute of Observational Sciences (ARIES),
Manora Peak, Nainital -- 263129, India\\
$^{2}$School of Studies in Physics \& Astrophysics, Pt. Ravishankar Shukla
University, Raipur-492 010 (CG), India  
}

\begin{document}

\date{Accepted ....... Received  ......; in original form ......}

\pagerange{\pageref{firstpage}--\pageref{lastpage}} \pubyear{2012}

\maketitle

\label{firstpage}

\begin{abstract}
Relative proper motions and cluster membership probabilities ($P_\mu$) have 
been derived for $\sim$ 2500 stars in the field of the open star cluster NGC~3766.
The cluster has been observed in $B$ and $V$ broadband filters at two epochs separated 
by $\sim$ 6 years using a wide-field imager mounted on the WFI@ESO{\bf 2.2}m 
telescope. All CCD frames were reduced using the astrometric techniques described 
in Anderson et al.\ (2006). The proper motion r.m.s. error for stars brighter than 
$V \sim 15$ mag is 2.0  mas~yr$^{-1}$ but it gradually increases up to  
$\sim$4 mas~yr$^{-1}$ at $V\sim20$ mag. Using proper motion data, membership probabilities 
have been derived for the stars in the region of the cluster. They indicate that three Be 
and one Ap stars are member of the cluster. The reddening $E(B-V)=0.22\pm0.05$ mag, 
a distance 2.5$\pm$0.5 kpc and an age of $\sim$ 20 Myr are derived using stars of 
$P_{\mu}>70\%$. Mass function slope $x=1.60\pm0.10$ is derived for the cluster and 
cluster was found to be dynamically relaxed. Finally, we provide positions, 
calibrated $B$ and $V$ magnitudes, relative proper motions and membership 
probabilities for the stars in the field of NGC 3766. We have produced a 
catalog that is electronically available to the astronomical community.

\end{abstract}


\section{Introduction}
The similar age of the stars in open clusters make them vital objects for 
the study of stellar evolution. However, for such studies, it is essential 
to separate the field stars from the cluster 
members. Proper motions (PMs) of the stars in the region of open clusters provide a 
unique possibility of getting membership information of the stars.
Proper motions in combination with radial velocities also provide valuable 
information on the kinematic parameters of the Galactic rotation curve and the 
distance to the Galactic center. They are also useful to study the formation and 
evolution of the open cluster system. 
   
The open cluster NGC 3766 ($\alpha_{2000}=11^{\rm h} 36^{\rm m} 14.^{\rm s}0$; 
$\delta_{2000}=-61^\circ 36\arcmin 30\arcsec$; $l=294\fdg12$; $b=-0\fdg03$) is 
located in the Carina complex of the Milky Way.
The cluster has been the target of many photometric studies (Ahmed (1962);
Yilmaz (1976); Shobbrook (1985), (1987); Moitinho et al. (1997); 
Tadross (2001); McSwain \& Gies (2005)) with the aim to determine the various 
physical parameters of the cluster. The most remarkable attribute of NGC 3766 
is the presence of largest number of Be stars yet known in any Galactic cluster. 
Among the 37 brightest stars, 12 show emission lines (Sher (1965); Schild (1970); 
Lloyd Evans (1980)). Fundamental parameters of B stars in NGC 3766 was studied 
by Aidelman et al. (2012) using spectroscopic data.

A proper motion study of NGC 3766 was made by Baumgardt et al. (2000) using eight 
stars taken from Hipparcos catalogue and derived the mean absolute proper motion 
$\mu_{\alpha}cos(\delta) = -7.21\pm0.34$ mas yr$^{-1}$ and $\mu_{\delta}=1.33\pm0.35$ mas yr$^{-1}$.

In spite of extensive photometric studies of this cluster, there is a lack of study 
providing proper motions and membership probabilities ($P_\mu$) of the stars in the wide-field 
region of open cluster NGC 3766. With the photometric data, membership information is 
desirable to minimize field star contamination. The archival wide-field multi-epoch CCD 
observations taken with the WFI@{\bf 2.2}m telescope provide an opportunity to derive 
precise PMs with a time-span of only a few years, and is deeper by several magnitudes 
than previous photographic surveys (Anderson et al. 2006; Yadav et al. 2008; 
Bellini et al. 2009; Sariya, Yadav \& Bellini 2012).

The main purpose of the present study is to estimate the fundamental parameters 
of the cluster using stars chosen according to their derived membership probabilities 
($P_\mu$). The $P_\mu$ of different sources in NGC 3766 is also discussed. 
We also derived luminosity function (LF), mass function (MF) and mass segregation 
of the cluster using stars of $P_\mu>$ 70\%. The PMs, $P_\mu$ and photometric $B$ and 
$V$ magnitudes are provided to the astronomical community for the follow-up studies. 

Observational data alongwith reduction procedures and determination of proper motions 
are described in the next Section. Colour-magnitude diagrams using proper motions are 
described in Sec.~\ref{cmd}. In Sect.~\ref{MP} we present cluster membership analysis. 
Sect.~\ref{par} is devoted to derivation of fundamental parameters of the cluster. 
Luminosity and mass function are described in Sec. \ref{sec:lf} while Sec. \ref{sec:ms} 
is devoted to the mass segregation study. In Sect.~\ref{cat} we describe our catalogue 
and finally, in Sect.~\ref{con} we present the conclusions of present study.


\section{Observational data and reductions}
\label{OBS}
CCD data for NGC 3766 were collected with the wide-field imager camera (WFI) 
mounted on {\bf 2.2}m ESO/MPI telescope at La Silla Chile. Using these archival 
data\footnote{http://archive.eso.org/eso/eso\_archive\_main.html.} of two epochs, 
proper motions were computed. The first epoch consists of four images in $B$ filter and 
three images in $V$ filter taken on 27$^{\rm th}$ Feb 2000, while second epoch have four 
images in $V$ filter taken on 9$^{\rm th}$ April 2006. The observational log  
is listed in Table~\ref{log}.

The WFI@{\bf 2.2}m consists of eight 2048$\times$4096 EEV CCDs with $0\farcs238$ 
pixel$^{-1}$ resulting in a total field-of-view 34$\arcmin \times 33\arcmin$. Images 
used in the analysis were taken between $1^{\prime\prime}.2-1^{\prime\prime}.5$ 
seeing condition and between 1.25-1.35 airmass. Long and short exposures were acquired 
to map the brighter as well as fainter stars of the cluster. 

\begin{table}
\caption{Description of the WFI@{\bf 2.2}m data sets used in this
  study. First epoch data was observed on Feb 27, 2000 while  second epoch 
data was observed on April 9, 2006.}
\label{log}
\begin{tabular}{ccccc}
\hline
\hline
Filters      &  Exposure Time & Seeing & Airmass&Limiting Mag. \\
&(in seconds)&&& \\
\hline
\multicolumn{4}{c}{(First epoch)} \\
$B$&2$\times$30; 2$\times$240&1$''$.5&1.25&$\sim$21     \\
$V/89$ &1$\times$30; 2$\times$240&1$''$.5&1.33&$\sim$20     \\
\multicolumn{4}{c}{(Second epoch)} \\
$V/89$ &4$\times$50&1$''$.2&1.35&$\sim$20        \\ 
\hline
\end{tabular}
\end{table}

\subsection{Astromeric and Photometric Reductions}

For the reduction of mosaic CCD images, we adopted the procedure described in 
Anderson et al. (2006, Paper~I). The procedure include de-biasing, flat-fielding and 
correction for cosmic rays. To get the position and flux of the stars in the image, 
we made an array of empirical Point Spread Functions (PSFs) for each image. An array of 
15 PSFs in each 2048$\times$4096 pixels chip (3 across and 5 high) as shown in Fig. 3 
of paper I was considered because PSFs changes significantly with position on the detector. 
In total, 120 PSFs have been considered for entire field of view (8192$\times$8192 pixels). 
These PSFs are saved in a look-up table on a very fine grid. To select suitable stars 
for the PSFs, an automatic code was developed (see Paper~I). An iterative process is designed 
to work from the brightest down to the faintest stars and find their precise 
position and instrumental flux for $B$ and $V$ exposures.

\begin{figure}
\vspace{-3cm}
\centering
\includegraphics[width=8.5cm]{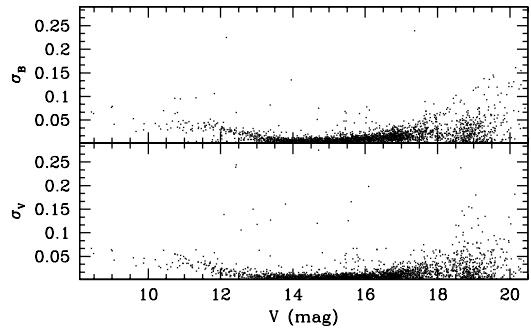}
\caption{Plot of rms of the residuals around the mean magnitudes
in $B$ and $V$ and as a function of $V$ magnitudes.}
\label{error_mag}
\end{figure}

In paper I, it is shown that WFI@{\bf 2.2}m has a large geometric distortion,
i.e. pixel scale is changing across the field of view (see Paper~I). To 
derive the correction for geometric distortion,we parametrized the 
distortion solution by a look-up table of corrections
for each chip that covered each 2048$\times$4096 pixel chip, sampling
every 256 pixels. This resulted in a 9$\times$17 element array of
corrections for each chip. At any given location on the  detector, a
bi-linear interpolation between the four closest grid points of the
look-up table provided the corrections for the
target point. The derived look-up table may have a lower accuracy on
the edges of a field because of the way the self-calibration frames
were dithered (see Paper~I). An additional source of uncertainty is
related to a possible instability of distortions for the $WFI@{\bf 2.2}$m
reported earlier. This prompted us to use the local-transformation
method to derive PMs. Detailed descriptions about the
distortion solution is given in Paper~I.

In the local transformation approach a small set of
local reference stars is selected around each target object. It is
advantageous to use pre-selected cluster members to form a local
reference frame because of a much lower intrinsic velocity dispersion
among the cluster members. Then, six-parameter linear transformations 
are used to transform the coordinates from one frame into another, 
taken at different epochs.

A general linear transformation from one frame $(x,y)$ into another 
$(u,v)$ has 6 parameters, and can be put into the following form:\\

$\left[ 
\begin{array}{c}
u\\v 
\end{array} 
\right]= \left[ \begin{array}{cc} A~~ B\\C~~ D \end{array} \right] \left[ \begin{array}{c} x-x_0\\y-y_0 \end{array} \right] + \left[ \begin{array}{c} u_0\\v_0 \end{array} \right]$ \\ 

The parameters $A, B, C$ and $D$ are the linear terms, and $x_0$, 
$y_0$, $u_0$ and $v_0$ are the constant terms. The equation look 
like there are 8 free parameters in the equations, but we can choose the zero-point 
in one frame arbitrarily, so only one of the two offsets is actually 
solved for. The easiest way to solve for the offsets is to adopt the 
centroid of the star list in the first frame as the offset for that 
frame, $(x_0, y_0)$. Then, the centroid in the other frame will be the 
least-squares solution for the other offset, ($u_0, v_0)$.

The residuals of this transformation are characterizing relative
PMs convolved with measurement errors. In essence, this is
a classical ``plate pair'' method but extended to all possible
combinations of the first and second epoch frames. Relative PM
of a target object is the average of all displacement measurements in
its local reference frame. The last step is to estimate the
measurement errors from intra-epoch observations where PMs
have a zero contribution. A complete description of all steps leading
to PMs is given in Paper~I.

\subsubsection{Photometric Calibration}

In order to transform instrumental $B$ and $V$ magnitudes into the
standard Johnson-Cousin system, we used the data published by Moitinho et al. (1997). 
Total 92 stars were used to calibrate $B$ and $V$ magnitudes.

We derived photometric zero-points and colour terms using the 
following transformation equations:\\

$ B_{\rm std} = B_{\rm ins} + C_b*(B_{\rm ins} - V_{\rm ins}) + Z_b $

$ V_{\rm std} = V_{\rm ins} + C_v*(B_{\rm ins} - V_{\rm ins}) + Z_v $ \\

where the subscript ``ins'' means instrumental magnitudes and ``std''
stands for secondary standard magnitudes. $C_b$ and $C_v$ are 
the colour terms while $Z_b$ and $Z_v$ are the global zero-points. 
The quadratic colour terms are negligible. The values of colour terms 
are 0.29 and 0.16, while the zero-points are 24.39 and 
23.79 for $B$ and $V$ filters respectively.

\begin{figure}
\centering
\includegraphics[width=8.5cm]{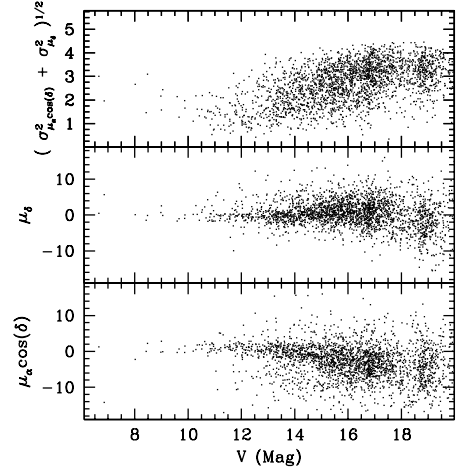}
\caption{Plots of proper motions and their rms error in mas~yr$^{-1}$
versus visual magnitude.}
\label{error_pm}
\end{figure}

In  Fig.~\ref{error_mag}  we show  the rms of the residuals around the mean 
magnitude for each filter as a function of the $V$ magnitude.
The photometric standard deviations have been computed from multiple 
observations, all reduced to the common photometric reference frame in the chosen bandpass.
The stars brighter than $\sim$ 12.0 mag in $V$ have higher dispersion because of
saturation. On average, photometric rms are better than $\sim$0.02 mag for 
stars between 12-18 mag and $\sim$ 0.04 mag between 18-20 mag in $V$ filter. 

\subsubsection{Astrometric Calibration}
\label{ac}

The $X, Y$ raw coordinates of each star in each frame were corrected for geometric 
distortion using the look-up table provided in Paper I, brought into a common reference frames
by means of six-parameter linear transformations, and averaged.
To transform the averaged $X$ and $Y$ coordinate into RA and Dec of
J2000, we used the online digitized sky ESO catalogue in skycat software, as an
absolute astrometric reference frame. Thanks to our accurate geometric-distortion solution and a
reasonable stability of the intra-chip positions, that it was possible to
apply a single plate model involving linear and quadratic terms and a
small but significant cubic term in each coordinate. This solution
also absorbs effects caused by differential refraction. 
The standard error of equatorial solution is
$\sim$ 30 mas in each coordinate.

\subsection{Determination of proper motions}
\label{ppm}

Proper motions were computed using $V$-filter images to minimize 
colour-dependent terms in our analysis. Moreover, our geometric-distortion 
solution provides the smallest residuals with $V$ images (Paper~I).
A total of three images for first epoch and four images for the second
epoch were used in proper motion determination.

First, we selected a sample of probable cluster members using the $V$
vs. $(B-V)$ CMD. Selected stars are located on the main-sequence in the magnitude 
range 8.0$\le V \le18.0$ mag. These stars define a local reference frame
to transform the positions of a given first-epoch image into positions
of a second-epoch image. By adopting only stars on cluster sequence 
having proper motion errors $<$3.0 mas~yr$^{-1}$, we assure that PMs are
measured relative to the bulk motion of the cluster. To minimize the effect of
uncorrected distortion residuals
we used the
local transformation approach based on the closest 25 reference stars
on the same CCD chip. No systematics larger than random errors are
visible close to the corners or edges of chips.

We iteratively removed some stars from the preliminary photometric
member list that had proper motions clearly inconsistent with cluster
membership, even though their colours placed them near the fiducial
cluster sequence. The distribution of proper motions rms error is 
presented in Fig. \ref{error_pm}. The precision of the proper
motion measurement is better than 3 mas yr$^{-1}$ upto 14 mag in 
$V$. They are gradually increasing upto 4 mas yr$^{-1}$ for $V=20$ mag.

\begin{figure}
\centering
\includegraphics[width=9.0cm,height=9.0cm]{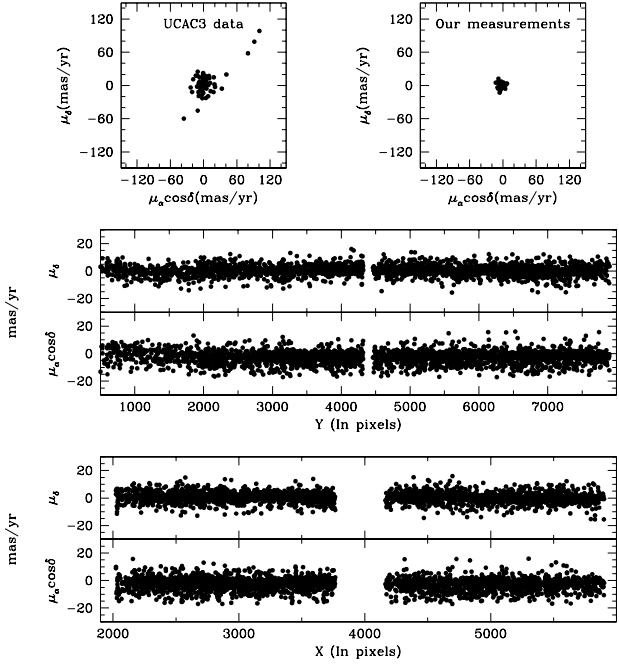}
\caption{{\it Top panels:} Vector-point diagrams of common stars
relative to the cluster mean motion for UCAC3 {\it (left)} and our catalogue {\it (right).
(Bottom panels:)} $\mu_{\alpha}Cos\delta$ and $\mu_{\delta}$ as a function of $X$ and $Y$ 
in $WFI$ pixels. Gaps in $X$ and $Y$ axis are the inter-chip space present in the mosaic CCD.} 
\label{ucac3}
\end{figure}

\subsubsection{Comparison with UCAC3 proper motion data}

UCAC3 catalogue (Zacharias et al. (2010)) provides the absolute proper motions 
of the stars in the region of NGC 3766. 
There are 94 common stars found to be brighter than $V=19.0$ mag.
To compare our proper motions with UCAC3, we changed the 
UCAC3 proper motions to relative proper motions. For this, we 
subtracted the mean absolute proper motion of the cluster
($\mu_{\alpha}cos{\delta} = - 3.4$ mas yr$^{-1}$, $\mu_{\delta}= -0.9$ mas yr$^{-1}$) 
from individual proper motion of stars. Fig. \ref{ucac3} shows 
the comparison of our PMs with those of the UCAC3 catalogue.
Top-left panel shows the vector-point diagrams (VPDs) of UCAC3 stars while top-right panel
shows the VPD of our measurements. A concentration of stars around (0,
0) mas yr$^{-1}$ is seen in both VPDs. Our proper motions distribution is
tighter than the UCAC3 distribution. This is because our data is
more precise than UCAC3 data.

In the lower panels of Fig. \ref{ucac3} we show the proper motions in $\mu_{\alpha}cos\delta$ 
and $\mu_{\delta}$ as a function of $X$ and $Y$ coordinates. There is no clear 
systematic trend seen in this diagram. 

\begin{figure*}
\centering
\includegraphics[width=\textwidth]{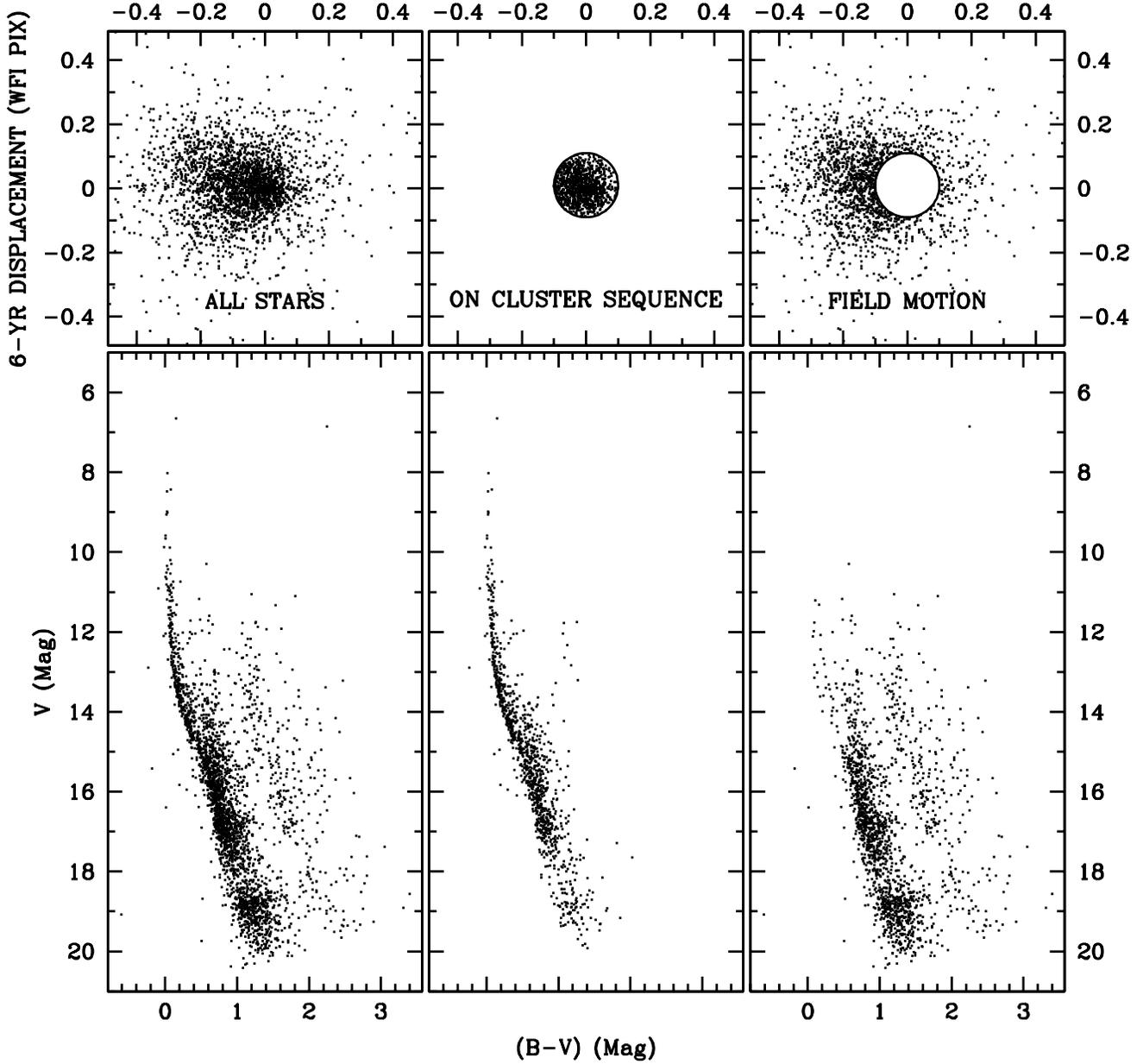}
\caption{
        {\em (Top panels)} Proper motion vector-point diagram.
        Zero point in VPD is the mean motion of cluster stars.
        {\em (Bottom panels)} Calibrated
$V$ vs. $(B-V)$ CMD.
        {\em (Left)} The entire sample;
        {\em (Center)} stars in VPD with proper motions within 4 mas~yr$^{-1}$
         around the cluster mean.
        {\em (right)} Probable background/foreground field stars in the direction
        of NGC~3766.
        All plots show only stars with proper motion
        $\sigma$ smaller than 3.0 mas~yr$^{-1}$ in each
        coordinate.  }
\label{cmd_inst}
\end{figure*}

%
\begin{figure*}
\centering
\includegraphics[width=\textwidth]{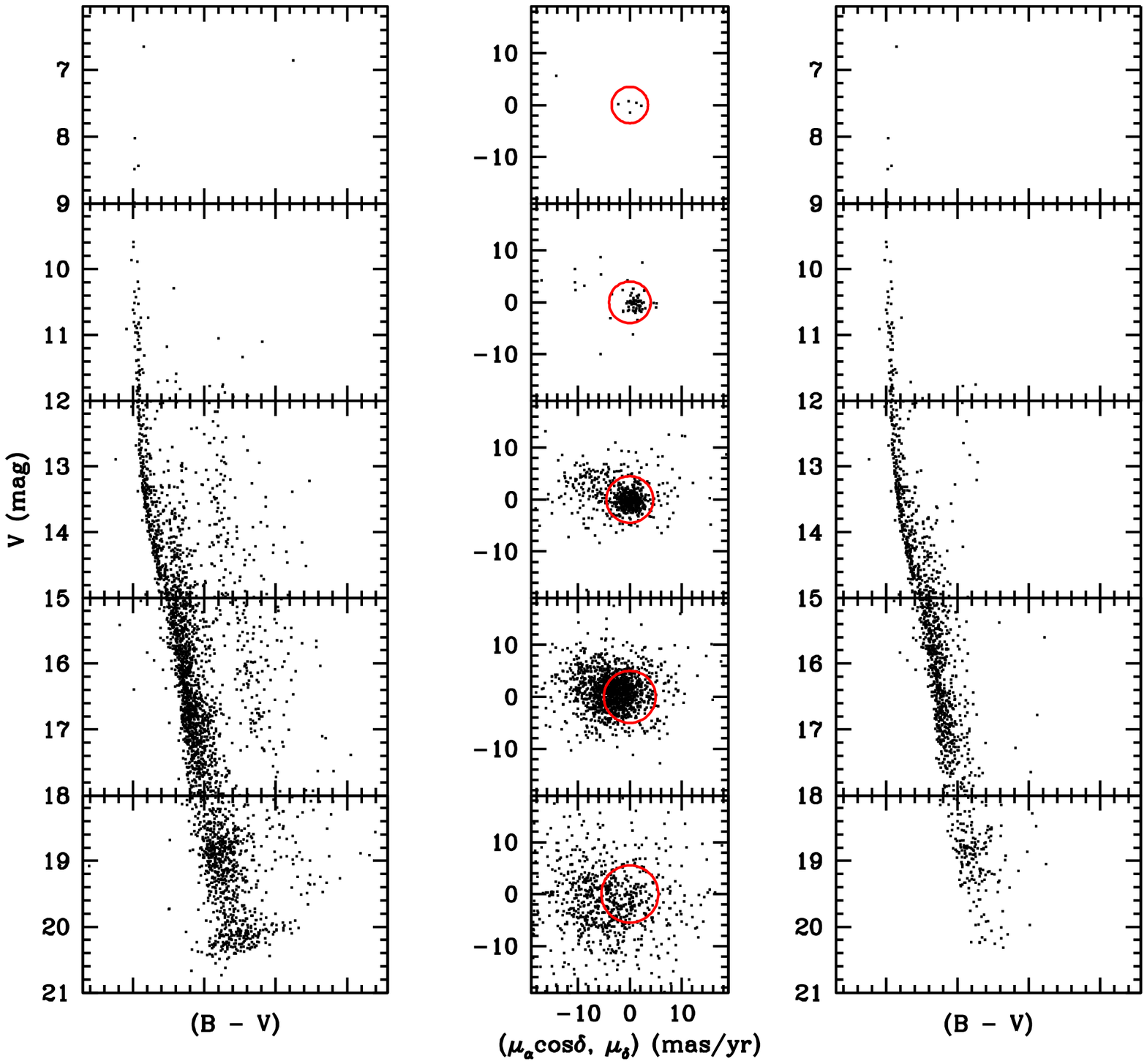}
\caption{
        {\em (Left:)} CMD for stars with proper motions.
        {\em (Middle:)} VPD for the same stars in
        corresponding magnitude intervals. A circle in each plot
        shows the adopted membership criterion.
        {\em (Right:)}  CMD for stars assumed to be cluster members.
        All plots show only stars with proper motion
        $\sigma$ smaller than 3 mas~yr$^{-1}$ in each
        coordinate.  }
\label{cmd_II}
\end{figure*}

\section{Cluster CMD decontamination}
\label{cmd}

PMs $\mu_{\alpha}cos\delta$ and $\mu_{\delta}$ are plotted as 
VPD in the top panels of Fig. \ref{cmd_inst}. Bottom 
panels show the corresponding $V$ vs. $(B-V)$ CMD. 
Left panels show all stars while middle and right panels show the probable cluster 
members and field stars. A circle of 4 mas yr$^{-1}$ around the cluster centroid 
in VPD of proper motions defines our membership criterion. The chosen radius 
is a compromise between losing cluster members with poor proper motions and 
including field stars sharing the cluster mean proper motion. The shape of 
cluster members PM dispersion is circular, providing that our PM measurements are 
not affected by any systematics. The CMD of most probable cluster 
members is shown in the lower middle panel. Main-sequence of the cluster is clearly 
separate out, which demonstrates the power of PMs derived in this study.
The lower right panel represents the CMD for field stars. Few cluster
members are also visible in this CMD because of their poorly determined 
proper motions. 

Fig. \ref{cmd_II} shows $(B-V), V$ CMD with VPD in various magnitude bins. We
divided the CMD into five magnitude bins and in each bin, we adopted 
different selection criteria to identify cluster members, which were
more stringent for stars with more reliable measurements from
data of high signal-to-noise ratio, and less restrictive for stars with
less precise measurements. These stars have a proper-motion
error of $\le$ 3.0 mas yr$^{-1}$. For each magnitude bin, we considered 
as cluster members those stars with a proper motion within the circle shown in
the middle column of Fig. \ref{cmd_II}. The right side of Fig. \ref{cmd_II} 
shows the color-magnitude diagram for stars assumed to be cluster members. 
This figure shows that the separation of brighter cluster members from the field stars
is clearly visible while fainter members are not clearly separated out.
The reason may be that proper motions for fainter stars are not determined accurately.
The available archive images are again not sufficiently deep to derive reliable
proper motions of fainter stars. The proper motions derived in
this paper are also not sufficiently accurate to study the internal motion 
of the cluster. 

\section{Membership Probabilities}
\label{MP}
%
The VPD for the proper motions derived in this study is shown 
in Fig.\ref{cmd_inst}. Present proper motion leads to the compact appearance 
of the cluster stars in VPD. Membership determination based on proper motions is 
useful for further astrophysical
studies in the region of cluster. The fundamental mathematical model set up by
Vasilevskis et al. (1958) and the technique based upon the maximum likelihood
principle developed by Sanders (1971) for membership determination have 
been continuously used and refined.

An improved method for membership determination of stars in clusters based
on proper motions with different observed precisions was developed by Stetson (1980)
and Zhao \& He (1990). Zhao \& Shao (1994) then added the
correlation coefficient of the field star distribution to the
set of parameters describing their distribution on the sky.

A tight clump at $\mu_{\alpha}$cos{$\delta$}=$\mu_{\delta}$=0.0 mas yr$^{-1}$
in Fig. \ref{cmd_inst} is representing the cluster stars and a broad distribution 
is seen for the field stars. 
Fig. \ref{cmd_II} indicates that the field stars centroid for panel 3 (middle column) 
is at a different location than for the last panel. Therefore, we consider two groups 
($V<=15$ mag and $V>15 $ mag) for membership determination. 
To determine the membership probability, we adopted the method
described in Balaguer-Nunez et al. (1998). This method has already
been used for $\omega$ Centauri (Bellini et al. (2009)) and by us for NGC 6809 (Sariya et al. 
(2012)). In this method
first we construct the frequency distribution of cluster stars
($\phi_c^{\nu}$) and field stars ($\phi_f^{\nu}$). The frequency
function for the $i^{\rm th}$ star of a cluster can be written as follows:\\

    $\phi_c^{\nu} =\frac{1}{2\pi\sqrt{{(\sigma_c^2 + \epsilon_{xi}^2 )} {(\sigma_c^2 + \epsilon_{yi}^2 )}}} exp\{{-\frac{1}{2}[\frac{(\mu_{xi} - \mu_{xc})^2}{\sigma_c^2 + \epsilon_{xi}^2 } + \frac{(\mu_{yi} - \mu_{yc})^2}{\sigma_c^2 + \epsilon_{yi}^2}] }\}$ \\

where $\mu_{xi}$ and $\mu_{yi}$ are the proper motions of the
$i^{\rm th}$ star while $\mu_{xc}$ and $\mu_{yc}$ are the cluster's proper motion
center. $\sigma_c$ is the intrinsic proper motion dispersion of
cluster member stars and ($\epsilon_{xi}, \epsilon_{yi}$) are the
observed errors in the proper motion components of $i^{th}$ star. The
frequency distribution for $i^{\rm th}$ field star is as follows:\\

$\phi_f^{\nu} =\frac{1}{2\pi\sqrt{(1-\gamma^2)}\sqrt{{(\sigma_{xf}^2 + \epsilon_{xi}^2 )} {(\sigma_{yf}^2 + \epsilon_{yi}^2 )}}} exp\{{-\frac{1}{2(1-\gamma^2)}[\frac{(\mu_{xi} - \mu_{xf})^2}{\sigma_{xf}^2 + \epsilon_{xi}^2}}$

$-\frac{2\gamma(\mu_{xi} - \mu_{xf})(\mu_{yi} - \mu_{yf})} {\sqrt{(\sigma_{xf}^2 + \epsilon_{xi}^2 ) (\sigma_{yf}^2 + \epsilon_{yi}^2 )}} + \frac{(\mu_{yi} - \mu_{yf})^2}{\sigma_{yf}^2 + \epsilon_{yi}^2}]\}$

where $\mu_{xi}$ and $\mu_{yi}$ are the proper motions of $i^{th}$ star while $\mu_{xf}$ and 
$\mu_{yf}$ are the field proper motion center. $\epsilon_{xi}$ and $\epsilon_{yi}$ are the observed 
errors in proper motions component and $\sigma_{xf}$ and $\sigma_{yf}$ are the field 
intrinsic proper motion dispersions and $\gamma$ is the correlation coefficient. It is 
calculated by

\begin{center}
$\gamma = \frac{(\mu_{xi} - \mu_{xf})(\mu_{yi} - \mu_{yf})}{\sigma_{xf}\sigma_{yf}}$
\end{center}

We did not consider the spatial distribution of the stars due to small observed 
field. To define the distribution function $\phi_c^\nu$ and $\phi_f^\nu$ we used 
stars having proper motion error better than 3 mas yr$^{-1}$. As expected, in the 
VPD, the center of cluster stars is found to be at $x_c$ = 0.0 mas yr$^{-1}$ and 
$y_c$ = 0.0 mas yr$^{-1}$. Our proper motion data set could not determine intrinsic 
proper motion dispersion ($\sigma_c$) for cluster stars. Assuming a distance of 2.5 kpc 
(present estimate) and radial velocity dispersion 1 kms$^{-1}$ for open clusters 
(Girard et al. (1989)), the expected dispersion in proper motions would be $\sim$ 
0.08 mas yr$^{-1}$. Therefore, we adopted $\sigma_c$ = 0.08 mas yr$^{-1}$. For
 field stars $V<= 15$ mag, we have: $x_f$ = $-$4.9 mas yr$^{-1}$, $y_f$ = 2.7 mas yr$^{-1}$,
 $\sigma_{xf}$ = 5.7 mas yr$^{-1}$  and $\sigma_{yf}$ = 4.0 mas yr$^{-1}$, while 
for $V>15$ mag: $x_f$ = $-$5.4 mas yr$^{-1}$, $y_f$ = 0.5 mas yr$^{-1}$ 
$\sigma_{xf}$ = 6.6 mas yr$^{-1}$  and $\sigma_{yf}$ = 5.3 mas yr$^{-1}$.  \\

The distribution of all the stars can be calculated as\\
~~~~~~~~~~~~~~~~~~~~~~~~~~~~~$\phi = (n_{c}~.~\phi_c^{\nu}) + (n_f~.~\phi_f^{\nu})$  \\

where $n_{c}$ and $n_{f}$ are the normalized number of stars for cluster and field 
($n_c + n_f = 1$). For the first group ($V<=15$ mag), $n_{c}=0.64$ and $n_{f}=0.36$ while 
for the second group ($V>15$ mag), $n_{c}=0.25$ and $n_{f}=0.75$. Therefore, the membership 
probability for $i^{th}$ star is \\

~~~~~~~~~~~~~~~~~~~~~~~~~~~~$P_{\mu}(i) = \frac{\phi_{c}(i)}{\phi(i)}$ \\

\begin{figure}
\vspace{-2.0cm}
\centering
\includegraphics[width=8.5cm]{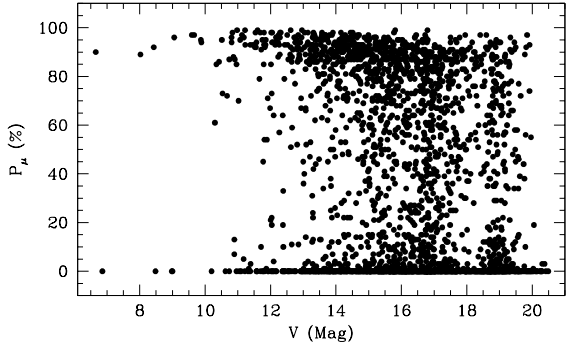}
\caption{Membership Probability $P_{\mu}$ as function  of the  $V$ magnitude,
for all the stars in our catalogue. At $V \sim 18$ mag and fainter, $P_{\mu}$
diminishes as a result of increasing errors in the PMs.}
\label{VvsMP}
\end{figure}

A good indicator of cluster and field separation is the membership probability. 
It is plotted as a function of magnitude in Fig. \ref{VvsMP}. As seen in this plot, 
high membership probability ($P_{\mu} > 90\%$) extend down to $V \sim 17$ mag. 
At fainter magnitudes the maximum probability gradually decreases.

\begin{figure}
\centering
\includegraphics[width=8.5cm]{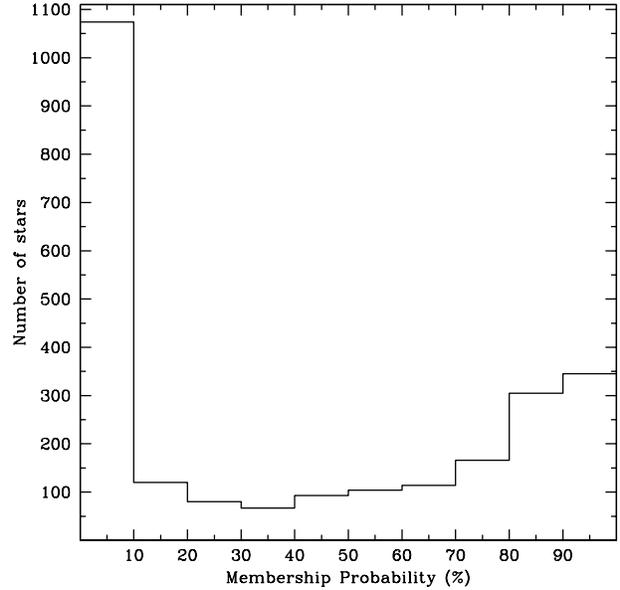}
\caption{Histograms of membership probabilities of all the stars in the region 
of the cluster NGC 3766}  
\label{mp_hist}
\end{figure}

Fig. \ref{mp_hist} shows the histogram of membership probabilities for 2468 stars 
in the cluster region. It shows a clear separation between cluster members
and field stars. This also indicates that the proper motion membership determination 
made using the method mentioned above is effective for the cluster NGC 3766. 
We find 818 stars with membership probabilities higher than 70\%. 

\subsection{Effectiveness of membership determination}

During the observations of the cluster we can not avoid the contamination 
by background and foreground objects. The effectiveness of our membership 
determination can be estimated using membership probabilities as described 
in Shao \& Zhao (1996). The effectiveness of membership determination is 
estimated as:\\

$E = 1-\frac{N\sum_{i=1}^N[P(i)\times(1-P(i))]}{\sum_{i=1}^N P(i)\sum_{i=1}^N (1-P(i))}$,\\

where the bigger E is, the more effective the membership determination is.

So we determine the effectiveness of membership determination as 0.65 for 
NGC 3766. It is shown in Fig. 3 of Shao \& Zhao (1996) that the effectiveness 
of membership determination of 43 open clusters are from 0.20 to 0.90 with peak 
value 0.55. Compared with the work by Shao \& Zhao (1996), we can see
that the effectiveness of membership determination for NGC 3766 is significantly 
high.

\begin{figure}
\centering
\includegraphics[width=8.5cm]{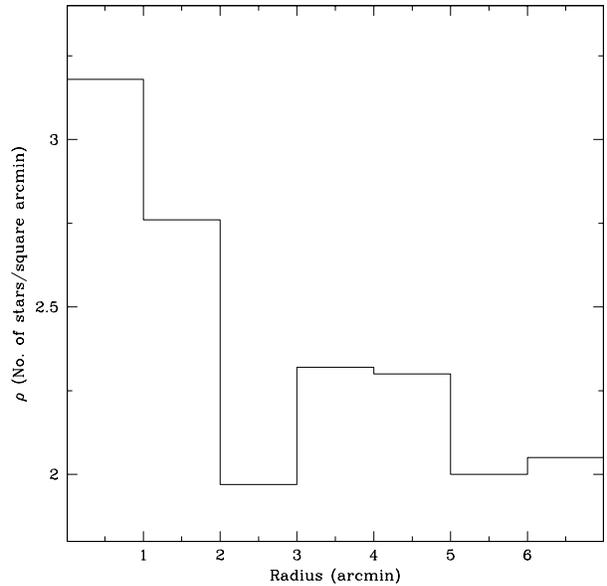}
\caption{Radial surface density distribution of the stars with $P_\mu>$ 70\% in NGC 3766 
region.}
\label{hist_rad}
\end{figure}

\subsection{Membership probability of published Be and Ap stars}

Present catalogue can be used to assign membership probability of various types of stars 
in this cluster. We found three Be stars and one Ap star in our catalogue according to the
 categorization in WEBDA. Membership probabilities of these stars are listed in Table 
\ref{vari}. The value of $P_{\mu}>=90\%$ for the listed stars indicate that they are 
member of the cluster.

\begin{table}  
\caption{Membership probabilities of Be and Ap stars in the region of NGC 3766.} 
\centering
\begin{tabular}{cccc}
\hline
$\alpha_{2000}$ & $\delta_{2000}$ & $P_{\mu}(\%)$ & Stars\\
\hline
$11^h36^m28^s.45$ &$ -61^{0}39^{\prime}54^{\prime\prime}.2$ &90 & Be\\
$11^h36^m31^s.60$ &$ -61^{0}34^{\prime}25^{\prime\prime}.1$ &90 & Be\\
$11^h36^m21^s.07$ &$ -61^{0}36^{\prime}57^{\prime\prime}.9$ &98 & Be\\
$11^h36^m14^s.11$ &$ -61^{0}37^{\prime}35^{\prime\prime}.2$ &94 & Ap\\

\hline
\label{vari}
\end{tabular}
\end{table}  

\section{Fundamental Parameters}
\label{par}

\subsection{Surface density distribution}

Fig. \ref{hist_rad} shows the radial density distribution of stars having 
membership probability $P_{\mu} > 70\%$. It is easily seen from the figure that 
the surface density distribution of the stars with radial distance ($r$) from 
the field center greater than $\sim$ 5 arcmin is quite flat, roughly being 
$\rho \sim$ 2 stars per square arcminutes, which is due to field stars 
with the same proper motion as NGC 3766. So we can reasonably choose 5 arcmin 
as the cluster radius. The cluster radius 4$^{\prime}.6$ derived by Moitinho et al. 
(1997) with stars $V<17$ mag is lower than the present estimate. If the distance 
of the cluster determined by us is adopted ($D=2500~ pc$), the corresponding linear 
radius is 3.6 pc. 

\subsection {Interstellar reddening, distance and age}

\begin{figure}
\centering
\includegraphics[width=8.5cm]{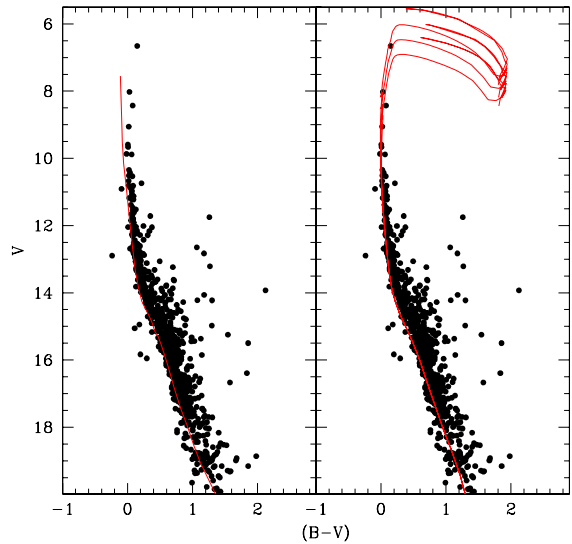}
\caption{$(B-V),V$ CMD of the cluster with the stars having $P_{\mu}>70\%$. 
In left panel, we have overplotted the ZAMS given by Schmidt-Kaler (1982) while right panel 
shows the isochrones fitting of different ages taken from Girardi et al. (2000).}
\label{dist}
\end{figure}

In order to determine the reddening and distance of
the cluster, the $(B - V), V$ CMD was
plotted in the left panel of Fig.~\ref{dist} for the stars $P_{\mu}>70\%$. 
The reddening and distance is estimated by fitting the ZAMS 
given by Schmidt-Kaler (1982) to the observational CMD. The best coincidence of ZAMS line 
with CMD of the cluster was achieved at the value of color 
excess $E(B-V) = 0.22\pm0.05$ mag. Our derived value of reddening 
agrees fairly well with the value $E(B-V) = 0.20\pm0.10$ mag estimated by 
Moitinho et al. (1997). The best fit ZAMS provides the value of 
apparent distance modulus $(v - M_v) = 12.75\pm0.05$ mag,
which corresponds to a distance of 2.5$\pm$0.5 kpc. For this cluster, our 
estimated value of distance is similar to the value 2.2$\pm$0.3 kpc
derived by Moitinho et al. (1997).

The age estimate of the cluster is based on visually fitting 
the theoretical isochrones described in Girardi et al. (2000). Isochrones 
of log(age) = 7.20, 7.30 and 7.40 are shown with continuous lines in the right 
panel of Fig. \ref{dist}.
The superposition of various isochrones on the CMD indicates that CMD is
represented well with an isochrone of about 20 Myr. The present age estimate 
is not too different from the value 21 Myr derived in the previous studies 
(Moitinho et al. (1997), Mermilliod \& Maeder (1986)).

\section{Luminosity and Mass function}
\label{sec:lf}
\begin{figure}
\centering
\includegraphics[width=8.5cm]{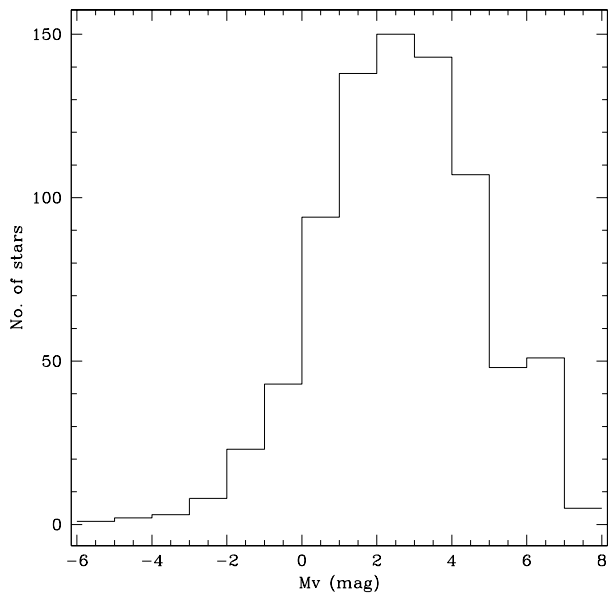}
\caption{The luminosity function of the stars in NGC 3766 region with $P_{\mu}$ 
higher than 70\%.}
\label{lf}
\end{figure}

In order to estimate the luminosity function (LF), histogram of the number of stars 
were constructed with 1.0 mag intervals in $M_v$ using the stars of $P_{\mu}>70\%$ 
and shown in Fig \ref{lf}. The interval of 1.0 mag was chosen so that there would be 
enough stars per bin for good statistics, as well as to provide a reasonable number 
of bins for the determination of mass function's slope. Fig \ref{lf} shows that the LF 
continues to rise to  $M_v\sim3$ mag and falls precipitously thereafter due to 
incompleteness in the photometry. With completeness becoming a problem after 
$M_v\sim3$ mag ($V$ $\sim$ 16 mag), the LF is complete upto $V\sim16$ mag. There 
is no apparent structure seen in LF to the level where the photometry is complete. 

\begin{figure}
\centering
\includegraphics[width=8.5cm]{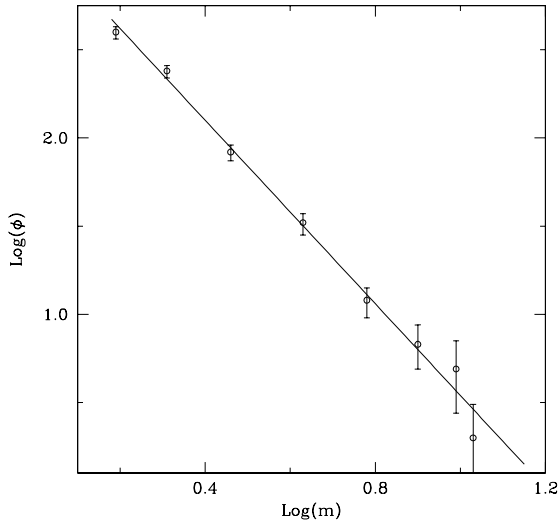}
\caption{Mass function for NGC 3766 derived using the stars with $P_{\mu}>70\%$ and brighter 
than 16 mag in $V$.}
\label{mf}
\end{figure}

Using the cluster's parameters derived in this analysis and theoretical models 
given by Girardi et al. (2000) we have converted LF to mass function (MF) and the 
resulting MF is shown in Fig \ref{mf}. The MF slope can be derived by using the 
relation log $dN/dM=-(1+x)$log$(M)$+constant, where $dN$ represents the number of 
stars in a mass bin $dM$ with central mass $M$ and $x$ is the slope of MF. This 
relation fit over the mass range $11.0\le M/M_\odot\le1.5$ ($8.0\le V \le16.0$ mag) 
gives a slope of -1.60$\pm$0.10 for the MF. Recently, Moitinho et al. (1997) 
derived the MF slope as -1.41$\pm$0.08 in the mass range $12.6\le M/M_\odot\le2.5$. 
Our derived value of MF slope is in agreement with Moitinho et al. (1997).

\section{Mass segregation}
\label{sec:ms}

\begin{figure}
\centering
\includegraphics[width=8.5cm]{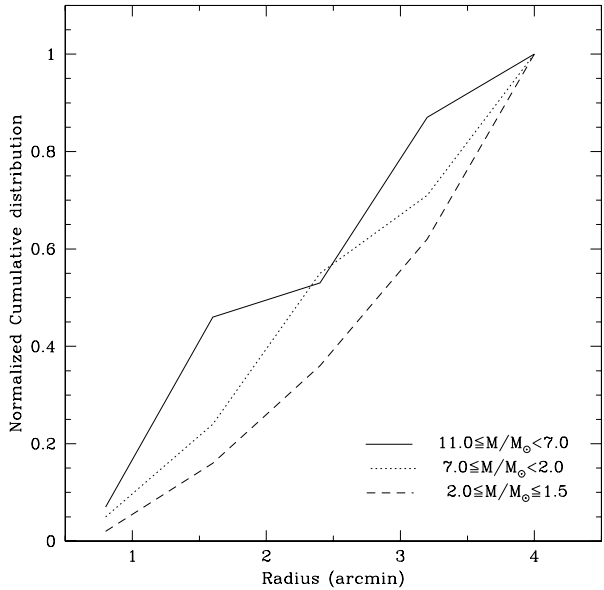}
\caption{The cumulative radial distribution of stars in various mass range.}
\label{ms}
\end{figure}

To study the cluster dynamical evolution and mass segregation, we selected the stars laying 
within the cluster radius and having $P_\mu>70\%$ with $V\le16$ mag. We subdivided the stars 
into three magnitude range, i.e. $8.0\le V<11.0$, $11.0\le V<14.0$ and $14.0\le V\le 16.0$ mag
corresponding to mass range $11.0\le M/M_\odot<7.0$, $7.0\le M/M_\odot<2.0$ and $2.0\le M/M_\odot\le1.5$. 
In Fig \ref{ms} we present cumulative radial stellar distribution of stars for different masses. 
An inspection of Fig \ref{ms} shows that the cluster exhibit a mass segregation effect. Massive 
stars seem to be centrally concentrated more than the low masses star. To check whether these 
mass distributions represent the same kind of distribution or not we perform the 
Kolmogorov-Smirnov (K-S) test. This indicates that mass segregation has taken place at a 
confidence level of 97\%. Further, it is important to investigate that whether existing mass 
segregation is caused by dynamical evolution or the imprint of star formation process.

One of the possible causes of mass segregation is the dynamical evolution of the cluster. Over 
the lifetime of star cluster, encounters between its member stars gradually lead to an increased 
degree of energy equipartition throughout the cluster. The most important result of this process 
is that the higher mass cluster members gradually sink towards the cluster center and in the 
process transfer their kinetic energy to the more numerous lower mass stellar component, thus 
leading to mass segregation. The time-scale on which a cluster will have lost all traces of its 
initial condition is well represented by its relaxation time $T_E$. It is given by

\begin{center}
\begin{displaymath}
\hspace{2.0cm}T_{E} = \frac {8.9 \times 10^{5} N^{1/2} R_{h}^{3/2}}{ <m>^{1/2}log(0.4N)}
\end{displaymath}
\end{center}

where $N$ is the number of cluster members, $R$$_{h}$ is the half-mass radius of the cluster and 
$<m>$ is the mean mass of the cluster stars (cf. Spitzer \& Hart (1971)). The value of $N$ and 
$<m>$ are 283 and 1.5 M$_{\odot}$ respectively. $R$$_{h}$ has 
been assumed as half of the cluster radius derived by us. Using the above relation we estimated the 
dynamical relaxation time $T_E =$ 10 Myr for NGC 3766. A comparison of cluster age with its 
relaxation time indicates that the relaxation time is smaller than the age of the cluster. 
Therefore, we can conclude that the cluster is dynamically relaxed.

\section {The catalog}
\label{cat}
The catalogue of proper motions with their corresponding uncertainties and membership 
probabilities for $\sim$ 2500 stars in the region of NGC 3766 is provided here. 
In the catalogue, Col. (1) contains the running number; Cols. (2) and (3) provide the 
J2000 equatorial coordinates, while Cols. (4) and (5) provide the pixel coordinates $x$ and $y$. 
Columns (6) to (9) represent relative proper motions and their standard errors. Cols. (10) to 
(13) gives photometric data i.e., $B$ and $V$ magnitudes and their 
corresponding errors. If photometry in a specific band is not available, a flag 
equal to 99.999 is set for the magnitude and 0.999 for the error. Cols. (14) gives 
the membership probability $P_{\mu}$. A sample of the data file is given in Table \ref{cata}.

\begin{table*}  
\caption{ First few lines of the electronically available catalogue. The stars listed in this table are found in all the $V$ images.} 
\centering
\tiny
\begin{tabular}{cccccccccccccc}
\hline\hline
ID & $\alpha_{2000}$ & $\delta_{2000}$ &x&y & $\mu_{\alpha}cos(\delta)$ &$ \sigma_{\mu_{\alpha}cos(\delta)}$ &$\mu_{\delta}$ & $\sigma_{\mu_{\delta}}$&B&$\sigma_B$&V&$\sigma_V$&$P_{\mu}$\\
(1)&(2)&(3)&(4)&(5)&(6)&(7)&(8)&(9)&(10)&(11)&(12)&(13)&(14) \\
&[h:m:s]&[d:m:s]&[pixel]&[pixel]&[mas/yr]&[mas/yr]&[mas/yr]&[mas/yr]&[mag]&[mag]&[mag]&[mag]&[$\%$]\\
\hline
     1& 11:35:48.75&   -61:51:46.7&  4501.195&   347.482&  -4.522&   1.269&  13.090&   2.856&  16.599&   0.013&  15.141&   0.009&    0\\
     2& 11:35:05.82&   -61:51:34.0&  5777.605&   399.354&   3.015&   1.428&   9.599&   2.618&  18.418&   0.029&  17.488&   0.011&    1\\
     3& 11:35:40.38&   -61:51:30.1&  4749.722&   417.288&  -3.729&   1.825&   3.134&   1.706&  16.168&   0.011&  15.499&   0.010&   23\\
     4& 11:35:47.94&   -61:51:28.3&  4525.247&   424.681&   5.196&   3.054&  -1.309&   2.975&  19.413&   0.001&  18.549&   0.062&   78\\
     5& 11:36:53.59&   -61:51:11.6&  2574.500&   490.833& -11.503&   2.261&  14.954&   1.785&  21.918&   0.128&  19.973&   0.133&    0\\
     -& ------     &   ------     &  ------  &   ------ & ------ &   -----&  ----- &   ---- &  ----  &   ---- &  ----  &   ---- &   -- \\  
\hline
\label{cata}
\end{tabular}
\end{table*}  

\section{Conclusions}
\label{con}
The purpose of this study is to provide a catalogue of precise proper motions and 
membership probability of stars in the field of open cluster NGC~3766.
We have obtained proper motions and astrometric membership
probabilities down to  $V\sim20$ mag in 15$\times$30 arcmin$^2$ area
around the open cluster NGC 3766. The fundamental parameters, LF, MF and mass segregation 
are derived using the stars of $P_{\mu}>70\%$. The linear radius of the cluster 
derived in the present study is 3.6 pc. The interstellar reddening $E(B-V)$ is 
found to be 0.22$\pm$0.05 mag. A distance of 2.5$\pm$0.5 kpc and an age of $\sim$ 
20 Myr have been derived for the cluster. The MF slope $x=1.60\pm0.10$ is estimated 
using the stars brighter than 16.0 mag in $V$. Present analysis indicates that NGC 3766 
is dynamically relaxed and one possible reason of this relaxation may be the dynamical 
evolution of the cluster.

Finally, we provide the membership probability for different types of stars, already 
found in the literature. Here, we also demonstrate that the CCD observations taken 
just 6 years apart can provide accurate proper motions and it can be used to separate 
the cluster members from field stars down to $V\sim20$ mag.\\ 

{\bf ACKNOWLEDGMENTS}\\
This work was based on observations with the MPG/ESO 2.2-m telescope located at La 
silla and Paranal Observatory, Chile, under DDT programs 164.O-0561(E) and 077.C-0188(B).
We thank the anonymous referee for a careful reading of the manuscript and for many useful 
comments. This research has made use of the VizieR catalogue access tool, CDS, Strasbourg, 
France and of the WEBDA open cluster database.

\end{document}